\documentclass[runningheads,a4paper]{llncs}

\usepackage{amssymb}
\usepackage{graphicx}

\newcommand{\be}{\begin{equation}}
\newcommand{\ee}{\end{equation}}
\newcommand{\bra}{\langle}
\newcommand{\ket}{\rangle}
\newcommand{\bea}{\begin{eqnarray}}
\newcommand{\eea}{\end{eqnarray}}
\newcommand{\dis}{\displaystyle}

\newcommand{\keywords}[1]{\par\addvspace\baselineskip
\noindent\keywordname\enspace\ignorespaces#1}

\begin{document}

\title{Markov Chain Monte Carlo on Asymmetric GARCH Model Using the Adaptive Construction Scheme}
\titlerunning{MCMC on Asymmetric GARCH Model}

\author{Tetsuya Takaishi}%

\institute{Hiroshima University of Economics,\\
731-0192  Hiroshima, Japan\\
\it E-mail: {takaishi@hiroshima-u.ac.jp}
}

\maketitle

\begin{abstract}
We perform Markov chain Monte Carlo simulations for 
a Bayesian inference of the GJR-GARCH model which is one of asymmetric GARCH models.
The adaptive construction scheme is 
used for the construction of the proposal density in the Metropolis-Hastings algorithm
and 
the parameters of the proposal density 
are determined adaptively by 
using the data sampled by the Markov chain Monte Carlo simulation.
We study the performance of the scheme 
with the artificial GJR-GARCH data.
We find that the adaptive construction scheme 
samples GJR-GARCH parameters effectively
and conclude that the Metropolis-Hastings algorithm with the adaptive construction scheme
is an efficient method to the Bayesian inference of the GJR-GARCH model. 
\end{abstract}

\section{Introduction}

Price returns of financial assets 
such as stock indexes, exchange rates 
show various interesting properties which can not be
derived from a simple assumption that the price returns follow the geometric Brownian motion.
Those properties are now classified as stylized facts\cite{Stanley,CONT}.
Typical examples of the stylized facts are
(i) fat-tailed distribution of return
(ii) volatility clustering
(iii) slow decay of the autocorrelation time of the absolute returns.

In empirical finance the volatility is an important quantity to measure risk. 
In order to forecast future volatility 
it is important to make models which mimic the properties of the volatility.
The most successful model is the Generalized Autoregressive Conditional Heteroscedasticity (GARCH)
model by Engle\cite{ARCH} and Bollerslev\cite{GARCH}, 
which can capture the property of volatility clustering and shows fat-tailed distributions.

In the original GARCH model\cite{ARCH,GARCH}, the volatility process is symmetric 
under positive and negative observations.
However the volatility often shows higher response against 
negative shocks, which was first observed by Black\cite{BLACK}.
This phenomenon is called the leverage effect.
In order to incorporate this asymmetry effect into the model,
some extended GARCH models have been proposed\cite{EGARCH,GJR,Ding,Zakoian,QGARCH1,QGARCH2}.
In this study we use the GJR-GARCH model by Glosten {\it et al.}\cite{GJR}
which introduces the asymmetry into the volatility process by an artificial indicator function.

In order to infer model parameters from financial data 
we employ Markov Chain Monte Carlo (MCMC) methods based on 
the Bayesian inference.
Since there is no unique way to implement the MCMC scheme
various MCMC methods for the GARCH models 
have been proposed\cite{Bauwens,Kim,Nakatsuma,Vrontos,WATANABE,ASAI,HMC,ARDIA}.
In a survey on the MCMC methods of the GARCH models\cite{ASAI} 
it is shown that Acceptance-Rejection/ Metropolis-Hastings  (AR/MH) algorithm
works better than other algorithms.
In the AR/MH algorithm the proposal density is assumed to be a multivariate Student's t-distribution and
the parameters to specify the distribution are estimated by the Maximum Likelihood (ML) technique.
Recently an alternative method to estimate those parameters without relying on the ML technique was proposed
\cite{ACS,ACS2,ACS3}.
In the method the parameters are determined by using the data generated by an MCMC method and
updated adaptively during the MCMC simulation.
We call this method "adaptive construction scheme".

The adaptive construction scheme was tested for GARCH and QGARCH models\cite{ACS,ACS2,ACS3} and
it is shown that the adaptive construction scheme can significantly reduce
the correlation between sampled data.
In this study we apply the adaptive construction scheme to
the GJR-GARCH model and study the efficiency of the adaptive construction scheme.

\section{GJR-GARCH Model}

The GJR-GARCH model\cite{GJR} is given by 
\be
y_t=\sigma_t \epsilon_t,
\ee
\be
\sigma_t^2  = \omega +  \alpha y_{t-1}^2 +\lambda \{1_{y_{t-1}<0}\} y_{t-1}^2+ \beta \sigma_{t-1}^2,
\label{eq:sigma}
\ee
where $\epsilon_t$ is an independent normal error $\sim N(0,1)$
and $y_t$ are observations.
The indicator function $1_{y_{t-1}<0}$ is 1 if $y_{t-1}<0$, and otherwise 0.
This indicator function introduces the asymmetry in the time series and
it generates higher volatilities after negative shocks than positive ones.

\section{Bayesian inference}
Using the Bayes' rule  
the posterior density $\pi(\theta|y)$  
with $n$ observations $y=(y_1,y_2,\dots,y_n)$ is given by
\be
\pi(\theta|y)\propto L(y|\theta) \pi(\theta),
\ee
where $L(y|\theta)$ is the likelihood function.
$\pi(\theta)$ is the prior density for $\theta$.
The functional form of $\pi(\theta)$ is not known a priori.  
In this study we assume that the prior density $\pi(\theta)$ is constant.

The likelihood function for the GJR-GARCH model is given by
\be
L(y|\theta)=\Pi_{i=1}^{n} \frac1{\sqrt{2\pi\sigma_t^2}}\exp\left.(-\frac{y_t^2}{2\sigma_t^2}\right.),
\ee
where $\theta=(\alpha,\beta,\omega,\lambda)$ stands for the GJR-GARCH parameters.
$\sigma_t^2$ is given by eq.(\ref{eq:sigma}).

Using $\pi(\theta|y)$ the GJR-GARCH parameters are inferred as the expectation values given by
\be
\bra {\bf \theta} \ket = \frac1{Z}\int {\bf \theta} \pi(\theta|y) d\theta,
\label{eq:int}
\ee
where $Z=\int \pi(\theta|y) d\theta$ is a normalization constant irrelevant
to MCMC estimations.

\subsection{Markov Chain Monte Carlo}
In general it is difficult to evaluate eq.(\ref{eq:int}) analytically. 
The MCMC technique gives a method to estimate eq.(\ref{eq:int}) numerically.
The basic procedure of the MCMC method is as follows.
First we sample $\theta$ drawn from a probability distribution
$\pi(\theta|y)$. 
Sampling is done by a technique which produces a Markov chain.  
After sampling  some data, 
we evaluate the expectation value as an average value over the sampled data $\theta^{(i)}$, 
\be
\bra {\bf \theta} \ket = \lim_{k \rightarrow \infty} \frac1k\sum_{i=1}^k \theta^{(i)},
\ee
where 
$k$ is the number of the sampled data.
The statistical error for $k$ independent data 
is proportional to $\frac1{\sqrt{k}}$.
In general, however, the data generated by the MCMC process are
correlated. As a result the statistical error will be proportional to $\sqrt{\frac{2\tau}{k}}$ 
where $\tau$ is the autocorrelation time among the sampled data.

\subsection{Metropolis-Hastings algorithm}

The MH algorithm\cite{MH} is a generalized version of the Metropolis algorithm\cite{METRO}. 
Let us consider to generate data $x$ from a probability distribution $P(x)$.
The MH algorithm consists of the following steps. 
First starting from $x$, 
we propose a candidate $x^{\prime}$ which is drawn from a certain probability distribution $g(x^{\prime}|x)$
which we call proposal density.    
Then we accept the candidate $x^{\prime}$ with a probability $P_{MH}(x,x^{\prime})$ 
as the next value of the Markov chain:
\be
P_{MH}(x,x^{\prime})= \min\left[1,\frac{P(x^{\prime})}{P(x)}\frac{g(x|x^\prime )}{g(x^{\prime}|x)}\right].
\label{eq:MH}
\ee
If $x^{\prime}$ is rejected we keep the previous value $x$.
We repeat these steps.

When the proposal density does not depend on the previous value, 
i.e. $g(x|x^\prime ) =g(x)$ we obtain
\be
P_{MH}(x,x^{\prime})= \min\left[1,\frac{P(x^{\prime})}{P(x)}\frac{g(x)}{g(x^{\prime})}\right].
\label{eq:MH2}
\ee

\section{Adaptive construction scheme}
The efficiency of the MH algorithm depends on how we choose the proposal density.
By choosing an adequate proposal density for the MH algorithm
one can reduce the correlation between the sampled data.  
The posterior density of GARCH parameters often resembles to a Gaussian-like shape. 
Thus one may choose a density similar to a Gaussian distribution as the proposal density.
Such attempts have been done by Mitsui, Watanabe\cite{WATANABE} and Asai\cite{ASAI}.
They used a multivariate Student's t-distribution in order to cover the tails of the posterior density and 
determined the parameters to specify the distribution by 
using the ML technique. 
Here we also use a multivariate Student's t-distribution 
but determine the parameters through MCMC simulations.

The ($p$-dimensional) multivariate Student's t-distribution is given by
\bea
g(\theta)& = & \frac{\Gamma((\nu+p)/2)/\Gamma(\nu/2)}{\det \Sigma^{1/2} (\nu\pi)^{p/2}} \nonumber \\
         &   & \times \left[1+\frac{(\theta-M)^t \Sigma^{-1}(\theta-M)}{\nu}\right]^{-(\nu+p)/2},
\label{eq:ST}
\eea
where $\theta$ and $M$ are column vectors,  
\be
\theta=\left[
\begin{array}{c}
\theta_1 \\
\theta_2 \\
\vdots \\
\theta_p
\end{array}
\right],
M=\left[
\begin{array}{c}
M_1 \\
M_2 \\
\vdots \\
M_p
\end{array}
\right],
\ee
and $M_i=E(\theta_i)$.
$\dis \Sigma$ is the covariance matrix defined as
\be
\frac{\nu\Sigma}{\nu-2}=E[(\theta-M)(\theta-M)^t].
\ee
$\nu$ is a parameter to tune the shape of Student's t-distribution. 
When $\nu \rightarrow \infty$ the Student's t-distribution goes to a Gaussian distribution.
At small $\nu$ Student's t-distribution has a fat-tail.
Since eq.(\ref{eq:ST}) is independent of the previous value of $\theta$, eq.(\ref{eq:MH2}) 
is used in the MH algorithm.

For the GJR-GARCH model,
$p=4$, $\dis \theta=(\theta_1,\theta_2,\theta_3,\theta_4)=(\alpha,\beta,\omega,\lambda)$,
and $\Sigma$ is a $4\times4$ matrix.
The unknown parameters in $M$ and $\Sigma$ are determined by using the data obtained from MCMC simulations.
First we make a short run by the Metropolis algorithm and accumulate some data.
Then we estimate $M$ and $\Sigma$. Note that there is no need to estimate $M$ and $\Sigma$ accurately. 
Second we perform an MH simulation with the proposal density of eq.(\ref{eq:ST}) with the estimated $M$ and $\Sigma$.
After accumulating more data, we recalculate $M$ and $\Sigma$, and update $M$ and $\Sigma$ of eq.(\ref{eq:ST}).
By doing this, we adaptively change the shape of eq.(\ref{eq:ST}) to fit the posterior density.

\section{Numerical Study}
\begin{figure}
\centering
\includegraphics[height=5cm]{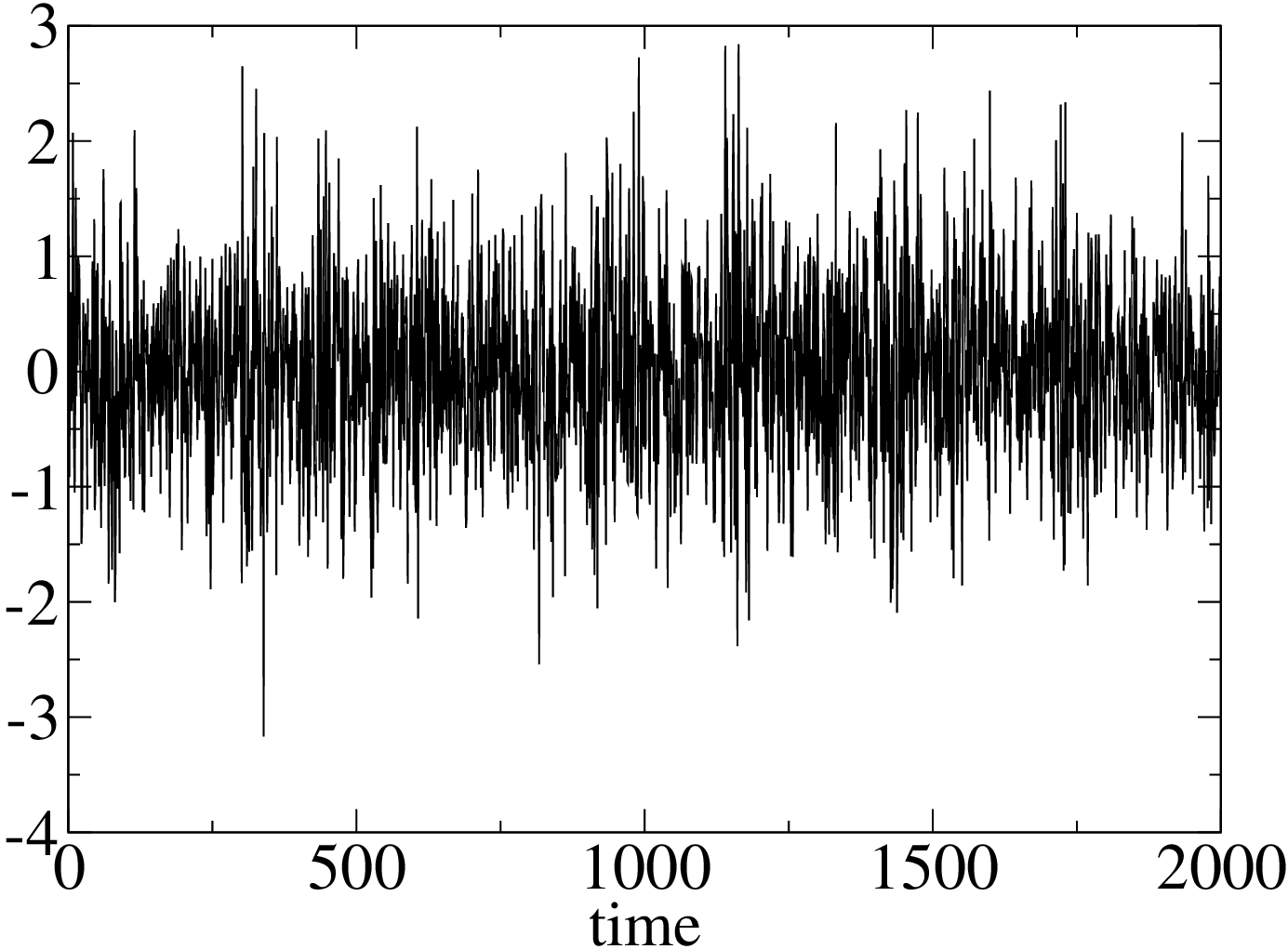}
\caption{
The artificial times series generated by the GJR-GARCH process with  $\alpha=0.03$, $\beta=0.85$, $\omega=0.05$ and $\lambda=0.1$.  }
\label{fig:DATA}
\end{figure}

In this section we study the performance of the adaptive construction scheme
by using artificial GJR-GARCH data generated with known parameters.
The GJR-GARCH parameters are set to $\alpha=0.03$, $\beta=0.85$, $\omega=0.05$ and $\lambda=0.1$.
We have generated 2000 data by the GJR-GARCH process with these parameters.
Fig.\ref{fig:DATA} shows the time series of the 2000 data.

The MCMC method with the adaptive construction scheme is implemented as follows. 
First we make a short run by a standard Metropolis algorithm.
The first 5000 data are discarded as burn-in process or thermalization.
Then we accumulate 1000 data for $M$ and $\Sigma$ estimations.
The estimated $M$ and $\Sigma$ are substituted to $g(\theta)$.
In this study we take $\nu=10$.
We re-start a run by the MH algorithm with $g(\theta)$. 
Every 1000 updates we re-calculate $M$ and $\Sigma$ and update $g(\theta)$.
We accumulate 100000 data for analysis.

For comparison we also use a standard Metropolis algorithm 
to infer the GJR-GARCH parameters 
and accumulate 100000 data for analysis.
In this study the Metropolis algorithm is implemented as follows.
We draw a candidate $\theta^\prime$ by 
adding a small random value $\delta \theta$ to the present value $\theta$:
\be
\theta^\prime = \theta + \delta \theta,
\ee
where $\dis \delta \theta= d(r-0.5)$.
$r$ is a uniform random number in $[0,1]$ and 
$d$ is a constant to tune the Metropolis acceptance.
We choose $d$ so that the acceptance becomes greater than $50\%$.

\begin{figure}
\vspace{3mm}
\centering
\includegraphics[height=4cm]{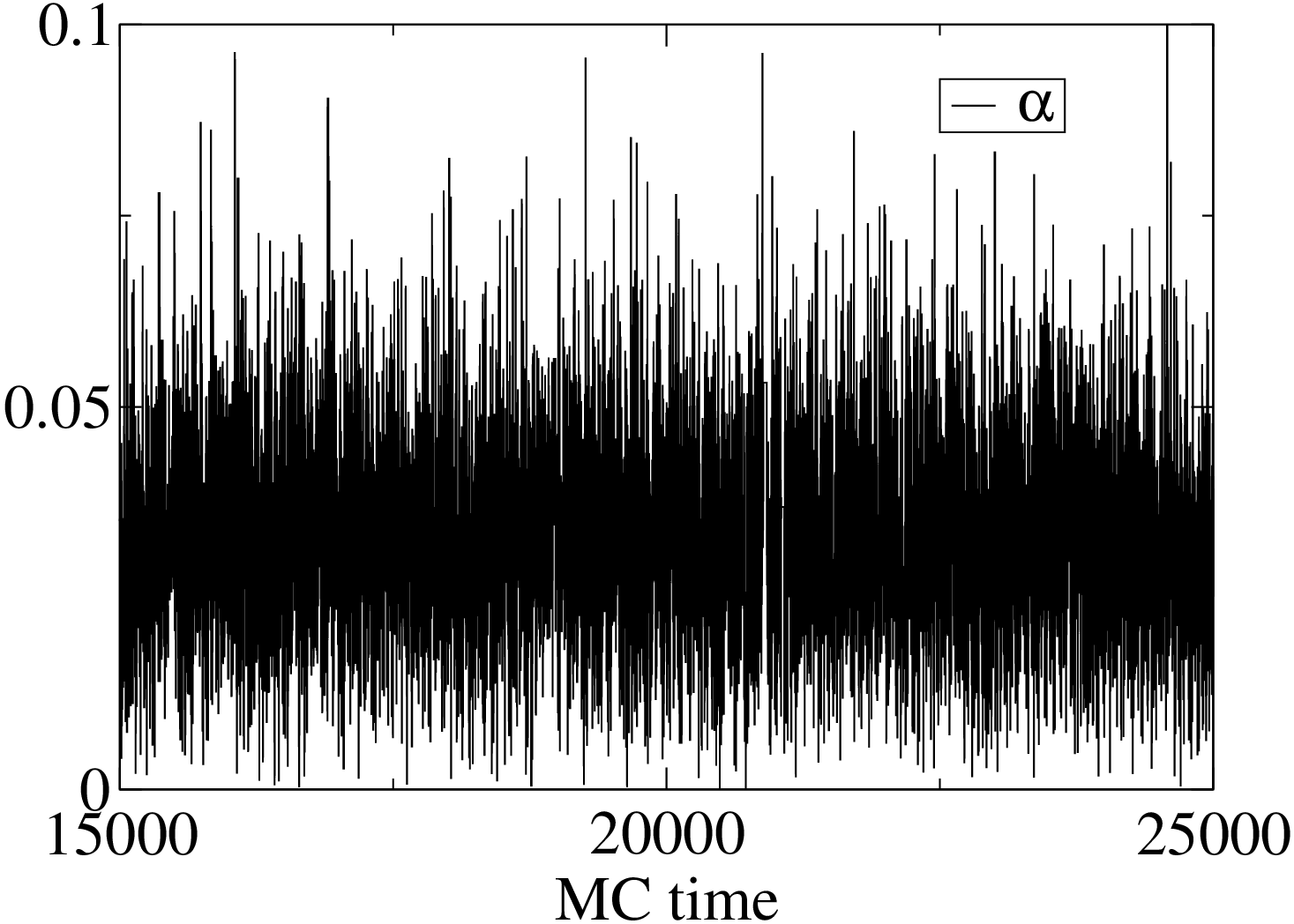}
\includegraphics[height=4cm]{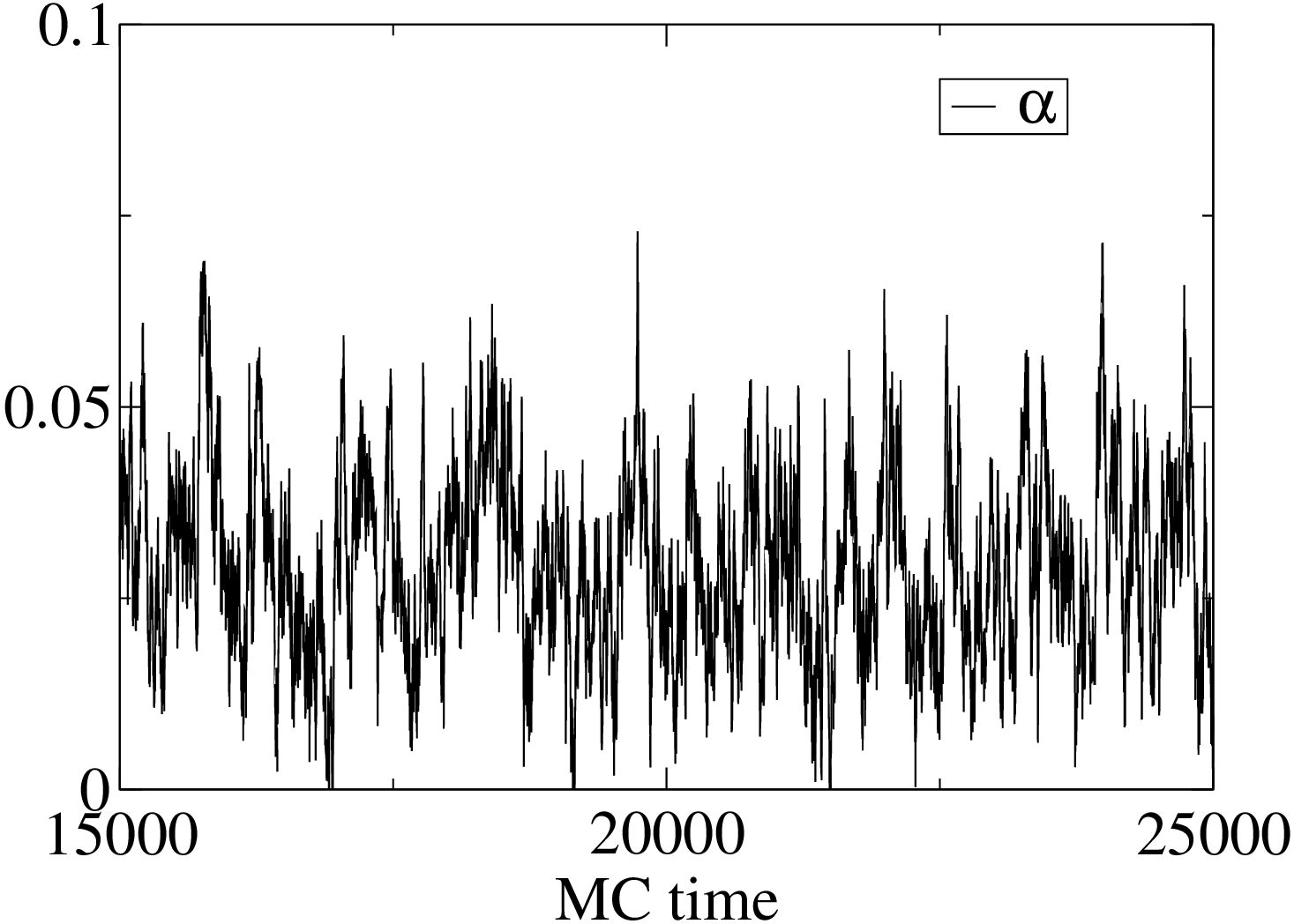}
\caption{
Monte Carlo history of $\alpha$ from the MH algorithm with the adaptive construction scheme(left) and
the Metropolis algorithm(right).
}
\label{fig:History}
\end{figure}

\begin{table}[h]
  \centering
  \caption{Results of GJR-GARCH parameters.
   SD and SE stand for standard deviation and statistical error respectively.}
  \label{tab:1}
  {\footnotesize
    \begin{tabular}{cllll}
      \hline
        & \multicolumn{1}{c}{$\alpha$} &
      \multicolumn{1}{c}{$\beta$} &
      \multicolumn{1}{c}{$\omega$}&
      \multicolumn{1}{c}{$\lambda$}         \\
      \hline
\hline
   true & 0.03 & 0.85 & 0.05 & 0.1 \\
\hline
   Adaptive  & 0.03285 & 0.85540  & 0.04522 & 0.08719 \\
   SD        & 0.0015  & 0.040    & 0.019   & 0.026   \\
   SE        & 0.00011 & 0.00025  & 0.00011 & 0.00066 \\
   $2\tau$   & $2.8 \pm 0.2$    & $3.3 \pm 0.7$ & $4.6 \pm 0.8$  & $2.6 \pm 0.2$  \\
\hline
   Metropolis  & 0.0323 & 0.855 & 0.0454 & 0.0895 \\
   SD          & 0.0015 & 0.038 & 0.018  & 0.026  \\
   SE          & 0.0007 & 0.004 & 0.0018 & 0.0015 \\
   $2\tau$      & $320\pm 100$  & $1050\pm 350$  & $990\pm 330$ & $350 \pm 110$ \\
      \hline
    \end{tabular}
  }
\end{table}

The results of the parameters inferred by the MH algorithm with the adaptive construction scheme and 
the Metropolis algorithm are summarized in Table 1.
We see that both algorithms well reproduce the values of the input parameters within 
the standard deviation. 
Furthermore the obtained values and the standard deviations from both algorithms coincide well each other. 
This observation is not surprising 
because both algorithms are performed using the same artificial data and thus
the posterior density is the same for both.  

Fig.\ref{fig:History} shows Monte Carlo time histories of the sampled $\alpha$ from the adaptive construction scheme (left)
and Metropolis algorithm(right). 
We see that the data sampled by the Metropolis algorithm are substantially correlated. 
The similar behavior is also seen for the sampled data for other parameters. 

The correlations between the data can be measured by the autocorrelation function (ACF). 
The ACF of certain successive data $\theta^{(i)}$ is defined by
\be
ACF(t) = \frac{\frac1N\sum_{j=1}^N(\theta^{(j)}- \bra\theta\ket )(\theta^{(j+t)}-\bra\theta\ket)}{\sigma^2_\theta},
\ee
where $\bra\theta\ket$ and $\sigma^2_\theta$ are the average value and the variance of $\theta$ respectively.

\begin{figure}
\vspace{5mm}
\centering
\includegraphics[height=4cm]{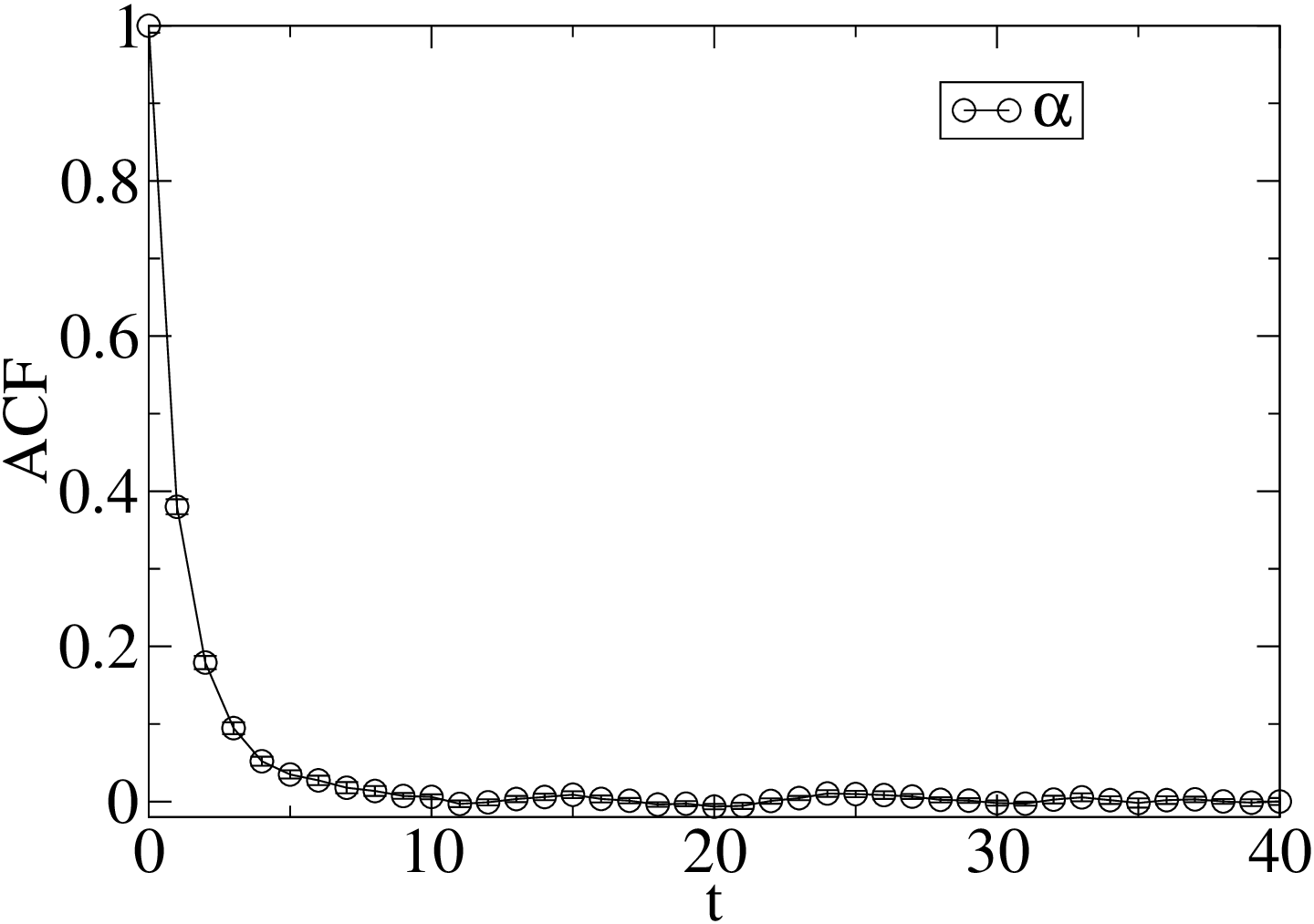}
\includegraphics[height=4cm]{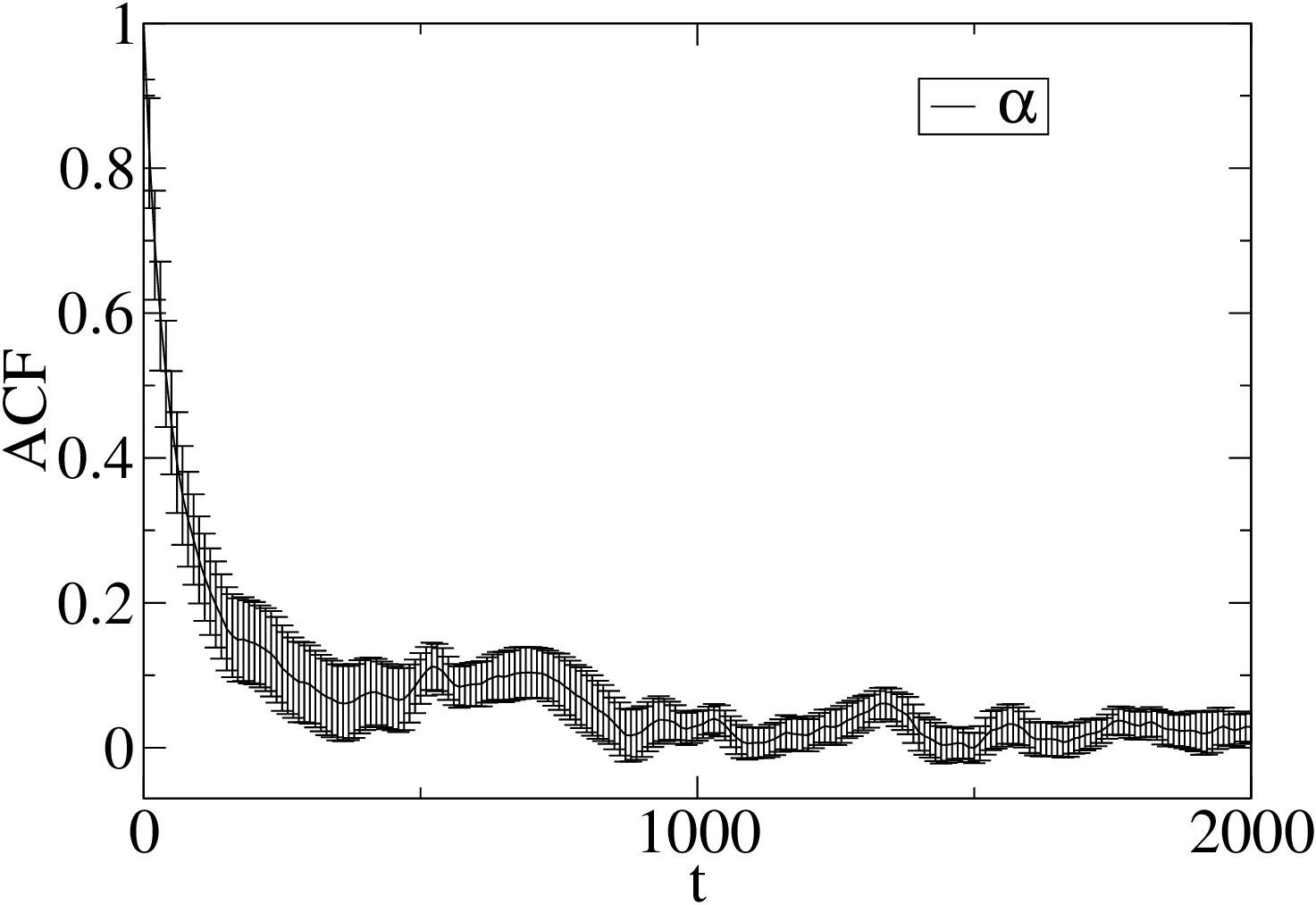}
\caption{
Autocorrelation function of $\alpha$ sampled by the adaptive construction scheme(left) and
the Metropolis algorithm(right).
}
\label{fig:ACF}
\end{figure}

Fig.\ref{fig:ACF} compares the ACF of $\alpha$ sampled from the adaptive construction scheme(left) 
and the Metropolis algorithm(rigth).
The ACF of the adaptive construction scheme decreases quickly as Monte Carlo time increases.
On the other hand the ACF of the Metropolis algorithm decreases very slowly, which indicates that
the correlation between the sampled data is very large.

To quantify the correlation between the data  
we calculate the autocorrelation time (ACT) $\tau$ defined by
\be
\tau = \frac12 +\sum_{i=1}^{\infty}ACF(i).
\ee
$2\tau$ is also called "inefficiency factor".
Results of $\tau$ are summarized in Table 1. 
We find that the ACT from the adaptive construction scheme
have much smaller $\tau$ than those from the Metropolis simulations. 
For instance $\tau$ of $\alpha$ parameter from the adaptive construction scheme is
decreased by a factor of 100 compared to that from the Metropolis algorithm. 
Furthermore the values of $\tau$ are about $2-3$ which are similar 
to the results of the AR/MH algorithm\cite{WATANABE,ASAI}.
These results  prove that the adaptive construction scheme is 
an efficient algorithm for sampling  de-correlated data.  
The differences in $\tau$ also explain that the statistical errors from the adaptive construction scheme
are much smaller than those from the Metropolis simulations.

\begin{figure}[t]
\vspace{5mm}
\centering
\includegraphics[height=4cm]{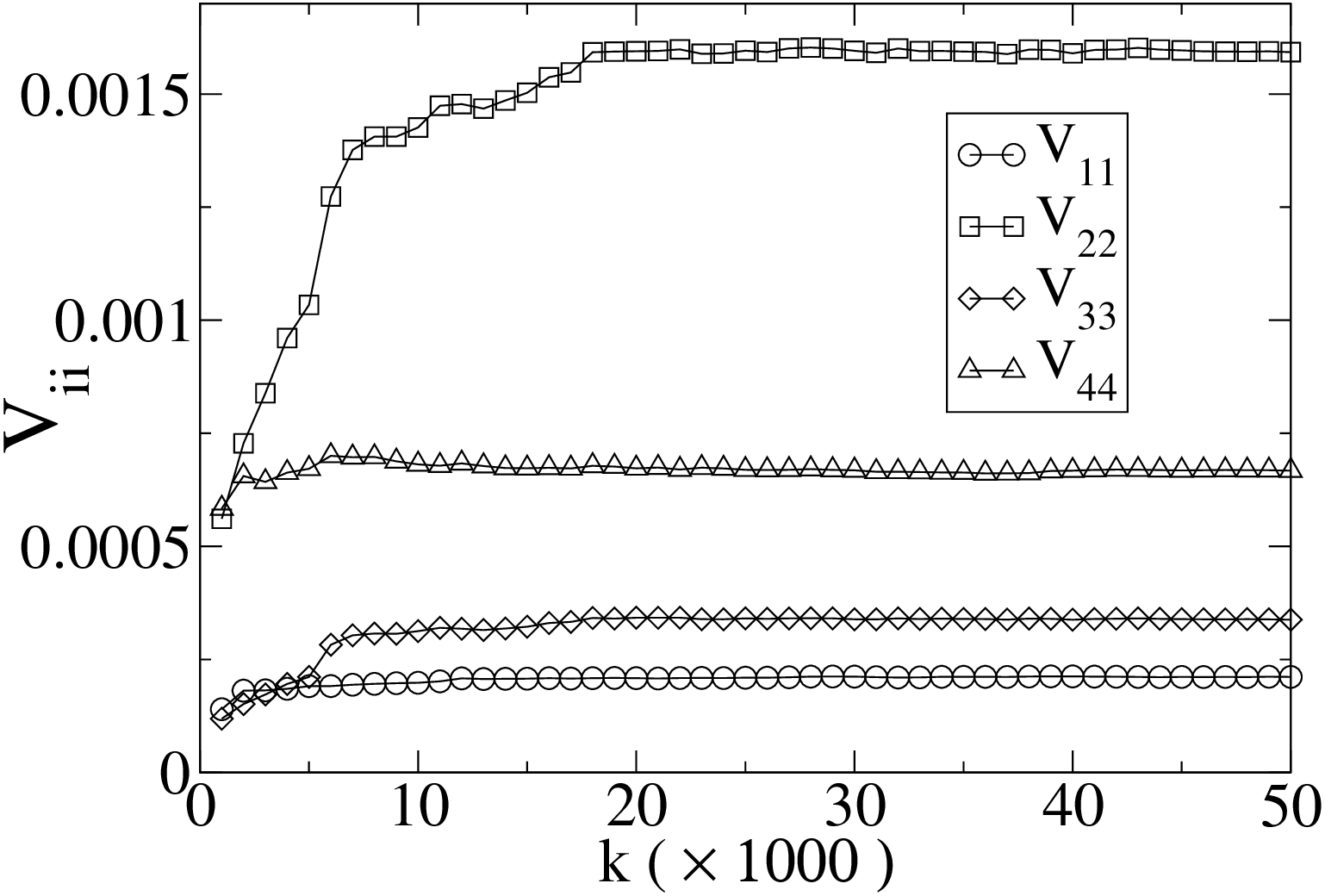}
\includegraphics[height=4cm]{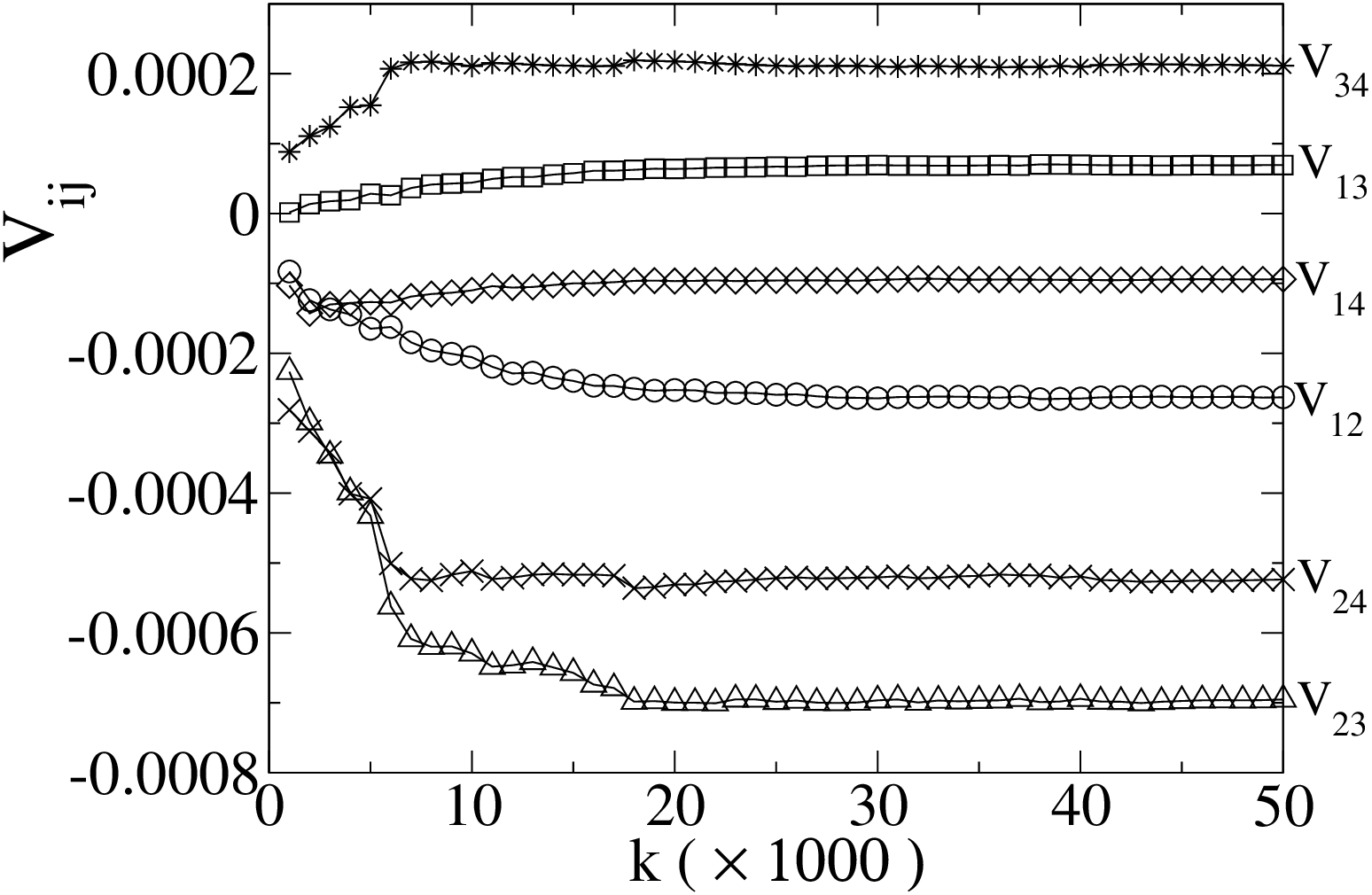}
\caption{
The diagonal(left) and off-diagonal(right) elements of $V$ as a function of the data size.
}
\label{fig:SIG}
\end{figure}

\begin{figure}
\vspace{5mm}
\centering
\includegraphics[height=5.0cm]{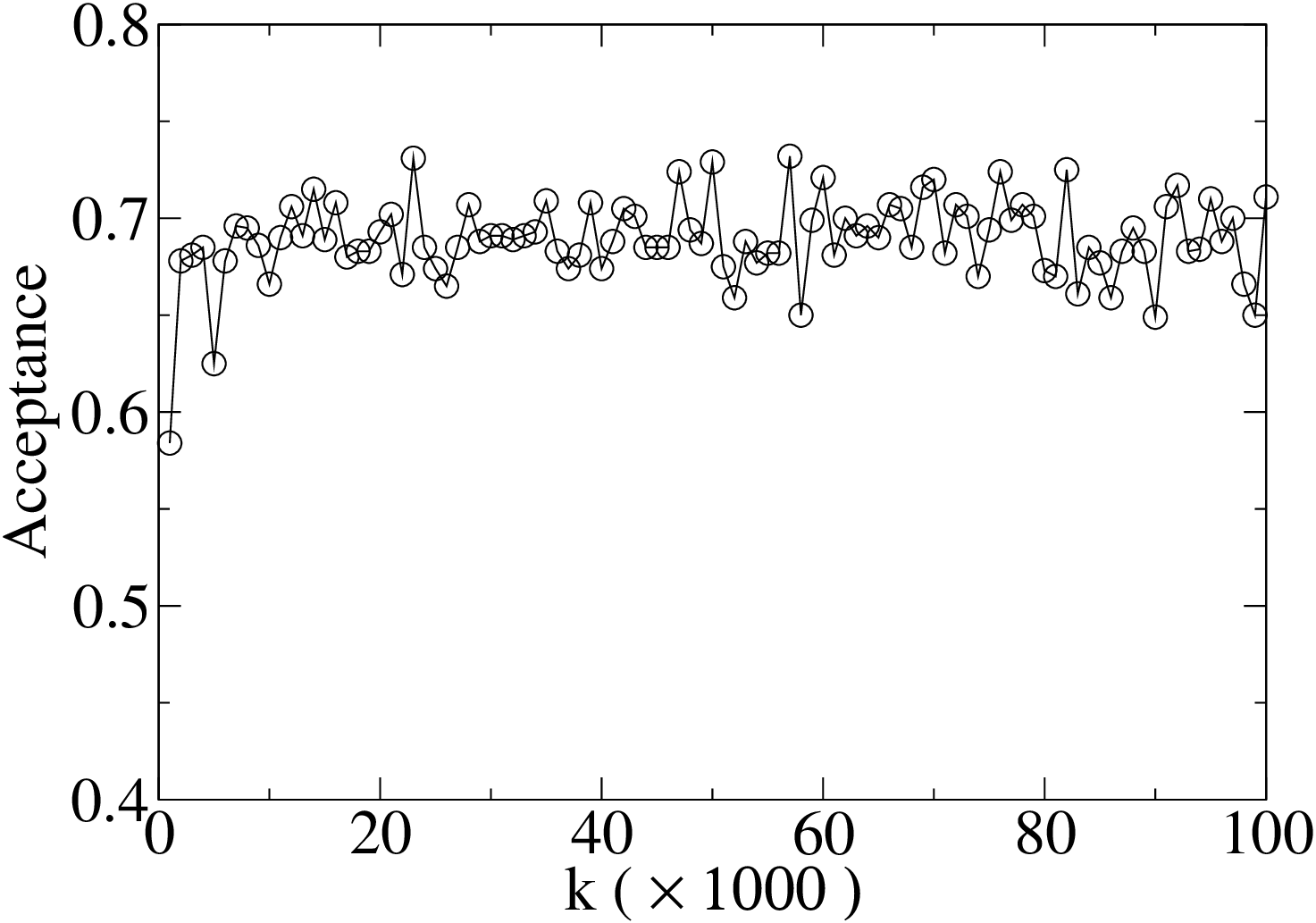}
\caption{
Acceptance at MH step with the adaptive proposal density.
}
\vspace{1mm}
\label{fig:ACC}
\end{figure}

Fig.\ref{fig:SIG} shows the convergence property of the diagonal and off-diagonal elements of the covariance matrix $V$.
Here $V$ is defined by $\dis V=E[(\theta-M)(\theta-M)^t]$.
We see that all the diagonal elements of $V$ quickly converge to certain values as 
the simulations are proceeded.

Fig.\ref{fig:ACC} shows the acceptance at the MH algorithm with the adaptive proposal density of eq.(\ref{eq:ST}).
Each acceptance is calculated every 1000 updates and the calculation of the acceptance is
based on the latest 1000 data.
Surprisingly
at the first stage of the simulation
the acceptance already reaches a plateaus of about $70\%$.
This indicates that the accuracy of the parameters of the proposal density are 
high enough for the MH algorithm already at the first stage.

\section{Conclusions}
We have performed the MCMC simulations of the Bayesian inference on the GJR-GARCH model
which is one of asymmetric GARCH models. 
The MCMC was implemented by the MH algorithm with the Student's t-distribution and 
the parameters of the Student's t-distribution were updated adaptively during the simulations.
The autocorrelation times of the data sampled by the adaptive construction scheme
are found to be very small. 
The obtained values of $\tau$ are similar to those of the AR/MH algorithm in \cite{WATANABE,ASAI}.
Thus the adaptive construction scheme has the similar efficiency with the AR/MH algorithm.
It is concluded that the adaptive construction scheme is also an efficient MCMC technique 
for the Bayesian inference of the GJR-GARCH model.
The previous studies on the GARCH and QGARCH models\cite{ACS,ACS2,ACS3} 
also reached the same conclusion.

It is also found that  
the acceptance of the MH algorithm quickly reaches a plateau of about $70\%$
at the beginning of the simulation.
This indicates that the parameters of the Student's t-distribution
for the MH algorithm
are calculated precisely enough by using the small data sampled at the 
beginning of the simulation.
This observation may suggest that 
one can stop updating the parameters at some point of the simulation
and use the same proposal density after that.

\section*{Acknowledgments}
The numerical calculations were carried out on SGI Altix3700 at the Institute of Statistical Mathematics
and on NEC SX8 at the Yukawa Institute for Theoretical Physics in Kyoto University. 
This study was carried out under the ISM Cooperative Use Registration (2009-ISM-CUR-0005).

\end{document}